\documentclass[preprint]{revtex4}
\input epsf
\usepackage{amsmath,amssymb}
\usepackage{graphicx}

\draft

\begin{document}
\titlepage
\title{$R^2$ corrections to the cosmological dynamics of inflation
 in the Palatini formulation}
\author{Xin-He Meng$^{1,2,3}$ \footnote{xhmeng@phys.nankai.edu.cn}
 \ \ Peng Wang$^1$ \footnote{pwang234@sohu.com}
} \affiliation{1.  Department of Physics, Nankai University,
Tianjin 300071, P.R.China \\2. Institute of Theoretical Physics,
CAS, Beijing 100080, P.R.China \\3. Department of Physics,
University of Arizona, Tucson, AZ 85721}

\begin{abstract}
We investigate the corrections to the inflationary cosmological
dynamics due to a $R^2$ term in the Palatini formulation which may
arise as quantum corrections to the effective Lagrangian in early
universe. We found that the standard Friedmann equation will not
be changed when the scalar field is in the potential energy
dominated era. However, in the kinetic energy dominated era, the
standard Friedmann equation will be modified and in the case of
closed and flat universe, the Modified Friedmann equation will
automatically require that the initial kinetic energy density of
the scalar field must be in sub-Planckian scale.
\end{abstract}

\maketitle

Inflation is an attractive paradigm for early cosmology (See e.g.
\cite{liddle} for a review). It assumes that the early universe
was filled with a scalar field, the "inflaton". When this scalar
field is in the slow-roll era, it can drive a de Sitter expansion.
Thus inflation can solve some cosmological puzzles such as the
horizon and flatness problems.

The main assumption of this paper is that initially the inflaton
field is kinetic energy dominated, i.e. $\frac{1}{2}\dot\phi^2\gg
V(\phi)$. This is a reasonable assumption in some inflationary
models. For example, considering Linde's  chaotic inflationary
scenario \cite{chaotic1}. This model indicates that the inflaton
field would take on a wide range of different values in different
regions of the universe, thus allows the inflation to occur under
quite general initial conditions. Specifically, Linde
\cite{chaotic2} has shown that in the context of General
Relativity, chaotic scenario can be realized even when the scalar
field obeys $\frac{1}{2}\dot\phi^2\gg V(\phi)$ and attains the
required condition for inflation in course of its evolution in FRW
universe. Furthermore, recently it has been suggested in
Ref.\cite{supp} that  in the chaotic inflation model, if the
kinetic dominated era is longer than the usual case so that the
inflaton field approaches the inflationary trajectory relatively
late, this may explain the suppressed lower multipoles in the CMB
anisotropies recently observed by WMAP \cite{wmap}. Thus it is
well-motivated to study the kinetic dominated era as the
pre-inflationary stage. In this era, it is reasonable to consider
the universe emerging from the Planck era with the scalar field
well displaced from any minimum, such that the typical energy is
of order the Planck energy. Thus quantum corrections to the
effective Lagrangian may play an important role. Starobinsky has
suggested that the effective Lagrangian in early universe should
include higher curvature terms such as a function of the Ricci
scalar $R$ \cite{Sta} (See e.g. \cite{odint} for a general review
of higher derivative gravity). Specifically, the $R^2$ corrections
to the chaotic inflation have been widely analyzed in the
literature, see e.g. Ref.\cite{Cardenas}. The corrections induced
by a $R^2$ term to early universe have also been investigated in
some other scenarios, see e.g. Ref.\cite{r2}. We will discuss the
$R^2$ term in this paper. However, it is conceivable that other
more complicated terms should display a qualitatively similar
behavior.

The starting point of the present work is that, generally, those
modified gravity models with a Lagrangian of the type $L(R)$ have
two \emph{inequivalent} formulations: the metric formulation
(second order formulation) and the Palatini formulation (first
order formulation) \cite{Volovich}. Recently, in
Refs.\cite{Vollick, Flanagan, Wang-R2, Wang1}, the Palatini
formulation is applied to the type of modified gravity models that
intend to explain the current cosmic acceleration without
introducing dark energy \cite{CDTT, Odintsov-mini, Odintsov-R2,
Dolgov}. There are many interesting features in the Palatini
formulation of those models, e.g. the absence of the instability
\cite{Wang1} appeared in the metric formulation \cite{Dolgov}
(However, this instability in metric formulation may be resolved
by a $R^2$ term \cite{Odintsov-R2} or conformal anomaly induced
terms \cite{Odintsov-mini}), the universality of the field
equations for vacuum or source with a constant trace of the
energy-momentum tensor \cite{Volovich}, among others.
Specifically, one interesting feature of the $R^2$ term in the
Palatini formulation is that a $R^2$ term alone cannot drive
inflation \cite{Wang-R2}, in opposite to the well-known result in
the metric formulation \cite{Sta, Cardenas}. Thus we expect that
in investigating the $R^2$ corrections to the inflation in the
Palatini formulation, we will find features that are different
from those that have been studied in Ref.\cite{Cardenas}. We have
found that this is indeed the case: the standard Friedmann
equation will not be changed when the scalar field is in the
potential energy dominated era; however, in the kinetic energy
dominated era, the standard Friedmann equation will be modified
and in the case of closed and flat universe, the Modified
Friedmann equation will automatically impose an upper bound on the
kinetic energy of the scalar field.

In general, when handled in Palatini formulation, one considers
the action to be a functional of the metric $\bar{g}_{\mu\nu}$ and
a connection $\hat{\bigtriangledown}_{\mu}$ which is another
independent variable besides the metric. The resulting modified
gravity action can be written as
\begin{equation}
S[\bar{g}_{\mu\nu}, \hat{\bigtriangledown}_{\mu}]=\int
d^4x\sqrt{-\bar{g}}[\frac{1}{2\kappa^2}(\hat R+\frac{\hat
R^2}{3\beta})-\frac{1}{2}(\partial_\mu\phi)^2-V(\phi)],\label{1}
\end{equation}
where we use the metric signature $\{-,+,+,+\}$, $\kappa^2=8\pi
G$, $\hat{R}_{\mu\nu}$ is the Ricci tensor of the connection
$\hat{\bigtriangledown}_{\mu}$, and
$\hat{R}=\bar{g}^{\mu\nu}\hat{R}_{\mu\nu}$. The $\beta$ is a
positive coupling constant with dimension $(mass)^2$,  and
$V(\phi)$ is the inflation potential.

The first question one would like to ask about the action (1) is:
what is the essential difference between the versions in the
Palatini formulation and in the metric formulation? The
differences will be more evident after we conformally transform
the action (1) to the Einstein frame following the general
framework of Flanagan \cite{Flanagan} (See the appendix for the
detailed derivation of this result):
\begin{eqnarray}
S[g_{\mu\nu}, \phi]&=&\int d^4x\sqrt{-g}[\frac{1}{2\kappa^2}
R-\frac{\beta/(2\kappa^2)}{2(4V+\beta/(2\kappa^2))}(\partial_\mu\phi)^2\\
\nonumber&+&\frac{1}{4(4V+\beta/(2\kappa^2))} (\partial_\mu\phi)^4
-(\frac{\beta/(2\kappa^2)}{4V+\beta/(2\kappa^2)})^2V(\phi)-\frac{\beta}{8\kappa^2}(\frac{4V}
{4V+\beta/(2\kappa^2)})^2]\label{1.5}
\end{eqnarray}
Thus we can see that when written in the Einstein frame, the
action (1) corresponds to the Einstein-Hilbert term plus a scalar
field with non-canonical kinetic term (This is actually a type of
higher derivative scalar field theory, see e.g. Ref.\cite{odint2}
for a review of the quantum properties of higher derivative scalar
field theory. Furthermore, in some models, these terms may be
responsible for instability of those models, see e.g.
Ref.\cite{odint3}). This is a type of k-inflation
\cite{k-inflation}. Thus our model can be viewed as a physical
example for the phenomenological model of k-inflation. In
Ref.\cite{Cardenas}, it is shown that when transformed to the
Einstein frame, the action (1) in metric formulation will
correspond to an Einstein-Hilbert term plus two scalar fields. So
there are essential differences in the Palatini and the metric
formulations of the same action. Thus it is not surprising that
the effects of the $R^2$ term will be rather different in the
metric and Palatini formulations. Specifically, in metric
formulation, the $R^2$ term and the scalar field $\phi$ may lead
to double inflation \cite{Cardenas}, i.e., two consecutive
inflationary stages separated by a power-law expansion. This will
not happen in the Palatini formulation. In this paper, we will
show that an interesting feature will appear in the Palatini
formulation that cannot be achieved in the metric formulation: the
$R^2$ term will automatically require that the initial kinetic
energy density of the scalar field must be in sub-Planckian scale.

One can proceed the analysis in Einstein frame and transform the
results to the physical Jordan frame. This is the method adopted
in the previous analysis of this model \cite{Cardenas}. Now there
is an much easier way. In Ref.\cite{Wang-R2}, we have derived the
Modified Friedmann (MF) equation of $R^2$ gravity for a flat
universe containing dust and radiation. To investigate the
corrections of $R^2$ term  to inflation, and since recently the
interest of inflation in closed universe is arousing (See e.g.
\cite{closed} and references therein), we will generalize this
derivation to arbitrarily perfect fluid with energy density $\rho$
and pressure $p$ and also to include closed geometry. We will see
that by using the MF equation the features of this model are quite
transparent.

Let us denote $L(\hat R)=\hat R+\frac{\hat R^2}{3\beta}$. Varying
the action (1) with respect to $\bar g_{\mu\nu}$ gives
\begin{equation}
L'(\hat R)\hat R_{\mu\nu}-\frac{1}{2}L(\hat R)\bar
g_{\mu\nu}=\kappa^2 T_{\mu\nu}\label{2.2}
\end{equation}
where a prime denotes differentiation with respect to $\hat R$ and
$T_{\mu\nu}$ is the energy-momentum tensor given by
\begin{equation}
T_{\mu\nu}=-\frac{2}{\sqrt{-\bar g}}\frac{\delta S_M}{\delta \bar
g^{\mu\nu}}\label{2.3}
\end{equation}
where $S_M$ is the matter action. For a universe filled with
perfect fluid, $T^{\mu}_{\nu}=\{-\rho,p,p,p\}$. Note that the
local conservation of energy momentum $\bar\nabla_\mu
T^{\mu\nu}=0$ is a result of the covariance of the action (1)
 and Neother theorem, thus it is independent of
the gravitational field equations (See e.g. Ref.\cite{wands}).
Then the energy conservation equation $\dot\rho+3H(\rho+p)=0$ is
unchanged.

When the early universe is dominated by the homogeneous scalar
field $\phi(t)$ with potential $V$, we have
\begin{equation}
\rho=\frac{1}{2}\dot\phi^2+V(\phi)\label{6}
\end{equation}
\begin{equation}
p=\frac{1}{2}\dot\phi^2-V(\phi)\label{7}
\end{equation}
The equation of motion for $\phi$ is of the standard form, which
can be obtained by varying the action (1) with respect to $\phi$
or from the fact that the energy conservation equation is
unchanged
\begin{equation}
\ddot\phi+3H\dot\phi+V'(\phi)=0\label{7.1}
\end{equation}

In the Palatini formulation, the connection is associated with
$\hat g_{\mu\nu}\equiv L'(\hat R)\bar g_{\mu\nu}$, which is known
from varying the action with respect to $\hat \Gamma
^{\lambda}_{\mu\nu}$. Thus the Christoffel symbol with respect to
$\hat g_{\mu\nu}$ is given in terms of the Christoffel symbol with
respect to $\bar g_{\mu\nu}$ by
\begin{equation}
\hat\Gamma ^{\lambda}_{\mu\nu}=\bar\Gamma
^{\lambda}_{\mu\nu}+\frac{1}{2L'}[2\delta
^{\lambda}_{(\mu}\partial _{\nu)}L'-\bar g_{\mu\nu}\bar
g^{\lambda\sigma}\partial _{\sigma}L']\label{Christoffel}
\end{equation}

Thus the Ricci curvature tensor is given by
\begin{eqnarray}
\hat R_{\mu\nu}=\bar R_{\mu\nu}+\frac{3}{2}(L')^{-2}\bar\nabla
_{\mu}L'\bar\nabla _{\nu}L' -(L')^{-1}\bar\nabla _{\mu}\bar\nabla
_{\nu}L'-\frac{1}{2}(L')^{-1}\bar g_{\mu\nu}\bar\nabla
_{\sigma}\bar\nabla ^{\sigma}L'\label{Ricci}
\end{eqnarray}
and
\begin{equation}
\hat R=\bar R-3(L')^{-1}\bar\nabla _{\mu}\bar\nabla ^{\mu}
L'+\frac{3}{2}(L')^{-2}\bar\nabla_{\mu}L'\bar\nabla^{\mu}L'\label{scalar}
\end{equation}
where $\bar R_{\mu\nu}$ is the Ricci tensor with respect to $\bar
g_{\mu\nu}$ and $\hat R=\bar g^{\mu\nu}\hat R_{\mu\nu}$. Note by
contracting Eq.(\ref{2.2}), we get:
\begin{equation}
L'(\hat R)\hat R-2L(\hat R)=\kappa^2 T\label{R(T)}
\end{equation}
Then we can solve $\hat R$ as a function of $T$ from
Eq.(\ref{R(T)}).
\begin{equation}
\hat R=-\kappa^2
T=\kappa^2(\rho-3p)=-\kappa^2(\dot\phi^2-4V(\phi))\label{RR}
\end{equation}
Thus equations (\ref{Ricci}) and (\ref{scalar}) do define the
Ricci tensor with respect to $\hat g_{\mu\nu}$.

We will consider the general Robertson-Walker metric (Note that
this is an ansatz for $\bar g_{\mu\nu}$ and is the result of the
assumed homogenization and isotropy of the universe, thus its form
is independent of the gravity theory):
\begin{equation}
ds^2=-dt^2+a(t)^2(\frac{dr^2}{1-kr^2}+r^2(d\theta^2+\sin^2\theta
d\phi^2))\label{metric}
\end{equation}
where $k$ is the spatial curvature and $k=-1,0,1$ correspond to
open, flat and closed universe respectively. The $a(t)$ is called
the scale factor of the universe.

From equations (\ref{metric}), and (\ref{Ricci}), we can get the
non-vanishing components of the Ricci tensor:
\begin{equation}
\hat
R_{00}=-3\frac{\ddot{a}}{a}+\frac{3}{2}(L')^{-2}(\partial_0{L'})^2-\frac{3}{2}(L')^{-1}\bar
\nabla_0\bar\nabla_0L'\label{R00}
\end{equation}
\begin{eqnarray}
\hat R_{ij}=[a\ddot{a}+2\dot{a}^2+2k+(L')^{-1}\bar\Gamma
^0_{ij}\partial_0L' +\frac{a^2}{2}(L')^{-1}\bar
\nabla_0\bar\nabla_0L']\delta_{ij}\label{ij}
\end{eqnarray}
where a dot denotes differentiation with respect to $t$.

Substituting equations (\ref{R00}) and (\ref{ij}) into the field
equations (\ref{2.2}), we can get
\begin{equation}
6H^2+3H(L')^{-1}\partial_0L'+\frac{3}{2}(L')^{-2}(\partial_0L')^2+6\frac{k}{a^2}=\frac{\kappa^2
(\rho+3p)+L}{L'}\label{2}
\end{equation}
where $H\equiv \dot{a}/a$ is the Hubble parameter, $\rho$ and $p$
are the total energy density and total pressure respectively.

Using the energy conservation equation $\dot\rho+3H(\rho+p)=0$, we
have
\begin{equation}
\partial_0L'=\frac{2\kappa^2}{\beta}(1-3c_s^2)(\rho+p)\label{4}
\end{equation}
where $c_s^2=dp/d\rho$ is the sound velocity.

Substituting Eq.(\ref{4}) into Eq.(\ref{2}) we can get the
Modified Friedmann (MF) equation of $R^2$ gravity in Palatini
formulation
\begin{eqnarray}
H^2&+&\frac{6k}{a^2[6-\frac{2\kappa^2(1-3c_s^2)\rho(1+\omega)}
{\beta+2\kappa^2\rho(1/3-\omega)}(1-\frac{\kappa^2(1-3c_s^2)\rho(1+\omega)}
{\beta+2\kappa^2\rho(1/3-\omega)})]}\\\nonumber&=&\frac{\frac{\kappa^2}{3}\rho[6+\frac{\kappa^2\rho}{\beta}(1-3\omega)^2]}
{[1+\frac{2\kappa^2\rho(1-3\omega)}{3\beta}][6-\frac{2\kappa^2(1-3c_s^2)\rho(1+\omega)}
{\beta+2\kappa^2\rho(1/3-\omega)}(1-\frac{\kappa^2(1-3c_s^2)\rho(1+\omega)}
{\beta+2\kappa^2\rho(1/3-\omega)})]}\label{5}
\end{eqnarray}
where $\omega=p/\rho$ is the effective equation of state. It is
interesting to note that when $\kappa^2\rho\ll\beta$, the MF
equation (18) reduces exactly to the standard Friedmann equation
$H^2+\frac{k}{a^2}=\frac{\kappa^2}{3}\rho$, independent of the
specifical form of $\rho$.

We will analyze the cosmological dynamics for the slow-roll and
kinetic dominated stages respectively.

First, when the scalar field is in the slow-roll region, we have
$\dot\phi^2\ll V(\phi)$, thus $\rho= V= -p$ and $\omega= -1$.
Substituting those to Eq.(18) we find that Eq.(18) reduces exactly
to the standard Friedmann equation
\begin{equation}
H^2+\frac{k}{a^2}=\frac{\kappa^2}{3}\rho=\frac{\kappa^2}{3}V(\phi)\label{8}
\end{equation}
Note that different from the remarks we make below Eq.(18), we do
not assume $\kappa^2\rho=\kappa^2V\ll\beta$ here. Thus, in
contrast to the case of metric formulation of the action (1)
\cite{Cardenas}, the appearance of a $R^2$ term will not affect
the standard inflation dynamics when the inflaton field is in the
slow-roll region.

Next we consider the kinetic energy dominated era, i.e.
$\dot\phi^2\gg V(\phi)$. In this case we have
$\rho=p=\frac{1}{2}\dot\phi^2$ and $\omega=c_s^2=1$ which
correspond to stiff matter.

Eq.(18) in this case reduces to
\begin{equation}
H^2+\frac{6k}{a^2[6+\frac{4\kappa^2\dot\phi^2(1+4\kappa^2\dot\phi^2/(3\beta))}
{\beta(1-2\kappa^2\dot\phi^2/(3\beta))^2}]}=\frac{\frac{\kappa^2}{6}\dot\phi^2(6+\frac{2\kappa^2}{\beta}\dot\phi^2)}
{(1-\frac{2\kappa^2\dot\phi^2}{3\beta})[6+\frac{4\kappa^2\dot\phi^2(1+4\kappa^2\dot\phi^2/(3\beta))}
{\beta(1-2\kappa^2\dot\phi^2/(3\beta))^2}]}\label{9}
\end{equation}

Now we can see that for a flat and closed universe model, i.e.
$k=0,1$, positivity of the right-hand side of this equation
requires that
\begin{equation}
\rho_{kin}=\frac{1}{2}\dot\phi^2<\frac{3\beta}{4\kappa^2}\equiv\rho^{max}_{kin}\label{10}
\end{equation}
Thus the kinetic energy of the scalar field is bounded from above.
Note that from the perspective of effective field theory, we
should have $\beta\sim M_p^2$. Thus, the bound (\ref{10}) says
that in presence of the $R^2$ term, we should require
$\rho_{kin}<M_p^4$. This bound is quite remarkable. In General
Relativity, there is no natural bound to prevent us from assuming
that initially $\rho_{kin}>M_p^4$ which is very problematic. Now
we can see that higher curvature terms such as the $R^2$ term may
automatically guarantee that the initial kinetic energy to be
sub-Planckian.

Next, to ensure that the MF equation does not prevent the
transition from kinetic energy dominated era to potential energy
dominated era \cite{chaotic2}, we consider the evolution in the
kinetic energy dominated era. We consider only the flat case for
simplicity. In this era, Eq.(\ref{7.1}) reduces to
\begin{equation}
\ddot\phi+3H\dot\phi=0\label{8.1}
\end{equation}
which can be integrated to give
\begin{equation}
\dot\phi^2=2\rho^0_{kin}(\frac{a_0}{a})^6\label{8.2}
\end{equation}
where $\rho^0_{kin}<\rho^{max}_{kin}$ is the initial value of the
kinetic density of the scalar field and $a_0$ is the initial value
of the scalar factor.

Substituting Eq.(\ref{8.2}) into Eq.(\ref{9}), we can get
\begin{equation}
H^2=\frac{C\beta}{6}(\frac{a_0}{a})^6\frac{[3+C(\frac{a_0}{a})^6][1-\frac{2C}{3}(\frac{a_0}{a})^6]}
{3-2C(\frac{a_0}{a})^6+4C^2(\frac{a_0}{a})^{12}}\label{11}
\end{equation}
where the constant $C\equiv3\rho^0_{kin}/(2\rho^{max}_{kin})<3/2$.

 Since we have $a>a_0$,
we can expand the right-hand side of Eq.(\ref{11}) up to the order
$(a_0/a)^{12}$, which reads
\begin{equation}
H^2=\frac{C\beta}{18}(\frac{a_0}{a})^6[3+C(\frac{a_0}{a})^6]\label{12}
\end{equation}

This equation can be integrated to obtain
\begin{equation}
a(t)=a_0\{\frac{3C\beta}{2}[t-t_0+\sqrt{\frac{2}{9\beta}+\frac{2}{3C\beta}}]^2-\frac{C}{3}\}^{\frac{1}{6}}\label{13}
\end{equation}
where $t_0$ is the initial time. Thus from Eq.(\ref{8.2}), we can
find that the scalar field evolves as
\begin{eqnarray}
\phi(t)=2\sqrt{\frac{\rho^0_{kin}}{3C\beta}}\{\ln[t-t_0+\sqrt{\frac{2}{9\beta}+\frac{2}{3C\beta}}
+\sqrt{(t-t_0+\sqrt{\frac{2}{9\beta}+\frac{2}{3C\beta}})^2-\frac{2}{9\beta}}]\\\nonumber
-\ln(\sqrt{\frac{2}{9\beta}+\frac{2}{3C\beta}}
+\sqrt{\frac{2}{3C\beta}})\}+\phi_0\label{14}
\end{eqnarray}
where $\phi_0$ is the initial value of the scalar field.

Thus, as in the case of General Relativity \cite{chaotic2}, for
sufficiently flat forms of the scalar potential, the
pre-inflationary stage ends very soon as the kinetic energy
decreases more quickly than the potential energy does, and then
the universe begins to inflate according to the standard scenario.

In conclusion, we have investigated the corrections to the
inflationary cosmological dynamics due to a $R^2$ term in the
Palatini formulation which may arise as quantum corrections to the
effective Lagrangian in early universe. We have argued that the
standard slow-roll stage will proceed as usual when the scalar
field is in the potential energy dominated era and hence the main
predictions of inflation model will not be changed. However, in
the kinetic energy dominated era that corresponds to the
pre-inflationary era, the standard Friedmann equation will be
modified and the Modified Friedmann equation will impose an upper
bound which states that the initial kinetic energy density of the
scalar field must be sub-Planck.

Note that in this paper we have only studied the two extremal
cases: kinetic energy dominated and potential energy dominated.
After the inflation ends, the inflaton field will enter the
coherent oscillation phase where the potential energy will be of
the same order as the kinetic energy. Considering the effects of
modified cosmological dynamics such as being induced by the $R^2$
term or in the braneworld scenario to the temperature evolution in
the reheating stage is an interesting thing to do, which may offer
us new bounds on the reheating temperature in order to constrain
the overproduction of gravitinos, black holes and many other
exotic objects. To study those effects, the main technical tool
one needs is the Modified Friedmann equation. Thus we can study
the $R^2$ correction to the reheating stage using the full MF
equation (18) derived in this paper. Due to its complicity, it can
only be studied numerically.

\textbf{Acknowledgements}

We would especially like to thank the referees' invaluable
comments and suggestions. We would also like to thank Profs.
D.Lyth, S.D.Odintsov and S.Nojiri for helpful discussions and
correspondences. This work is partly supported by China Doctoral
Foundation of National Education Ministry and ICSC-World
Laboratory Scholarship.

\textbf{Appendix:}

The general framework of the correspondence between the Jordan and
Einstein frames for Palatini formulation has been developed in
Ref.\cite{Flanagan}. Adopting this framework, it can be shown that
action (1) is conformally equivalent to
\begin{equation}
S[g_{\mu\nu}, \alpha, \phi]=\int d^4x\sqrt{-g}[\frac{1}{2\kappa^2}
R-\widetilde{V}(\alpha)-\frac{1}{2}e^{2\alpha}(\partial_\mu\phi)^2-e^{4\alpha}V(\phi)]\label{1.1}
\end{equation}
where
\begin{equation}
g_{\mu\nu}=e^{-2\alpha}\bar{g}_{\mu\nu}\label{1.2}
\end{equation}
\begin{equation}
\widetilde{V}(\alpha)=\frac{\beta}{8\kappa^2}(1-e^{2\alpha})^2\label{1.3}
\end{equation}
Note that the scalar field $\alpha$ has no kinetic term, thus it
is not a dynamical degree of freedom. Its equation of motion can
be solved to give
\begin{equation}
e^{2\alpha}=\frac{-(\partial_\mu\phi)^2+\beta/(2\kappa^2)}{4V+\beta/(2\kappa^2)}\label{1.4}
\end{equation}

Substituting this into Eq.(\ref{1.1}), we can integrate the
$\alpha$ field out from the action and the resulting action is
just the form of Eq.(2)
\begin{eqnarray}
S[g_{\mu\nu}, \phi]&=&\int d^4x\sqrt{-g}[\frac{1}{2\kappa^2}
R-\frac{\beta/(2\kappa^2)}{2(4V+\beta/(2\kappa^2))}(\partial_\mu\phi)^2\\
\nonumber&+&\frac{1}{4(4V+\beta/(2\kappa^2))} (\partial_\mu\phi)^4
-(\frac{\beta/(2\kappa^2)}{4V+\beta/(2\kappa^2)})^2V(\phi)-\frac{\beta}{8\kappa^2}(\frac{4V}
{4V+\beta/(2\kappa^2)})^2]
\end{eqnarray}
Thus we can see that when written in the Einstein frame, the
action (1) corresponds to the Einstein-Hilbert term plus a scalar
field with non-canonical kinetic term.

\end{document}